\documentclass[preprint,preprintnumbers,amsmath,amssymb,nofootinbib,aps]{revtex4}
\pdfoutput=1
\usepackage{epsfig}

\usepackage{graphicx}
\usepackage{dcolumn}
\usepackage{bm}
\usepackage{amsmath}
\usepackage{latexsym}
\usepackage{color}
\usepackage{hyperref}
\usepackage{mathrsfs}

\newcommand{\be}{\begin{equation}}
\newcommand{\ee}{\end{equation}}

\begin{document}


\title{Acceleration in Friedmann cosmology with torsion}

\author{S. H. Pereira$^{1}$} \email{s.pereira@unesp.br}
\author{R. de C. Lima$^{1}$} \email{rodrigo.lima@feg.unesp.br}
\author{J. F. Jesus$^{2,1}$}\email{jf.jesus@unesp.br}
\author{R. F. L. Holanda$^{3}$} \email{holandarfl@fisica.ufrn.br}

\affiliation{$^1$Universidade Estadual Paulista (UNESP), Faculdade de Engenharia de Guaratinguet\'a, Departamento de F\'isica e Qu\'imica, Guaratinguet\'a, SP, 12516-410, Brazil
\\$^2$Universidade Estadual Paulista (UNESP),  C\^ampus Experimental de Itapeva, Itapeva, SP, 18409-010, Brazil
\\$^3$Departamento de F\'isica Te\'orica e Experimental, Universidade Federal do Rio Grande do Norte, Natal, RN, 59300-000, Brazil
}


\def\zt{\mbox{$z_t$}}

\begin{abstract}
A Friedmann like cosmological model in Einstein-Cartan framework is studied when the torsion function is assumed to be proportional to a single $\phi(t)$ function coming just from the spin vector contribution of ordinary matter. By analysing four different types of torsion function written in terms of one, two and three free parameters, we found that a model with $\phi(t)=- \alpha H(t) \big({\rho_{m}(t)}/{\rho_{0c}}\big)^n$ is totally compatible with recent cosmological data, where $\alpha$ and $n$ are free parameters to be constrained from observations, $\rho_m$ is the matter energy density and $\rho_{0c}$ the critical density. The recent accelerated phase of expansion of the universe is correctly reproduced by the contribution coming from torsion function, with a deceleration parameter indicating a transition redshift of about $0.65$. 

\end{abstract}

\maketitle



\section{Introduction}
\label{sec:introd}
A more complete understanding of general relativity with the presence of matter can be obtained when one consider that the intrinsic angular momentum of fermionic particles (spin) promotes torsion effects in space-time. This can be achieved with the presence of asymmetric affine connection in the construction of a manifold, introducing the torsion of spacetime and therefore allowing emerge of new geometric degrees of freedom in the system. Thus, matter becomes responsible for being a source of torsion, enriching studies in cosmological scenarios, with more general prescriptions. An example is based on well-established studies of the Einstein-Cartan-Kibble-Sciama (ECKS) gravitational theory. This theory allows to describe in a more complete way the invariance of local gauge in relation to the group of Poincar\`e \cite{Sciama:1964wt,Hehl:1971qi,Hehl:1976kj,Shapiro:2001rz}, being very useful in studies of condensate of particles with half-integer spin and averaged as a spin fluid \cite{Hehl:1974cn,Sabbata90,Sabbata94} besides scenarios with an effective ultraviolet cutoff in quantum field theory for fermions \cite{Poplawski:2009su}. Even though there is no observational evidence to ponder the existence of torsion in spacetime, some suggestions for experimental tests involving spacetime studies with non-zero torsion for gravity can be found in \cite{Kostelecky:2007kx,March:2011ry,Hehl:2013qga}. One of the major problems in finding this evidence is associated with the fact that effects of torsion become considerable mainly at high density and energies. {There are other cosmological scenarios that have interesting consequences generated by torsion corrections. In \cite{Gonzalez-Espinoza:2019ajd} the torsion effects generated by scalar fields contributes to explain inflation. In addition, non-minimal couplings with torsion effects have been studied to understand gravitational waves \cite{Valdivia:2017sat,Izaurieta:2019dix}.}
  
However, Friedmann-Robertson-Walker (FRW) cosmological scenarios also can be addressed in presence of torsion. In particular, the very tiny value of the cosmological constant or dark energy needed to accelerate the universe could be mimicked due to contribution of the torsion. Moreover, the high symmetry of FRW spacetime preserves the symmetry associated to Ricci curvature tensor, which implies that the corresponding Einstein tensor and energy-momentum tensor also preserve a symmetric form. Such construction is very well motivated and discussed in \cite{Kranas2019}, which we recommend for further details. The whole effect of torsion due to spin of matter may be associated to a single scalar function, depending only on time. Such approach was also adopted in \cite{Olmo:2011uz,Capozziello:2008kb,Jimenez:2015fva,Poplawski:2010kb}. {A dynamical system approach with weak torsion field was done recently in \cite{barrow2019}}. A review on Friedmann cosmological models in Einstein-Cartan framework can be found in \cite{Medina2019}.  The kinematics of cosmological spacetimes with nonzero torsion in the context of classical Einstein-Cartan gravity is given by \cite{Pasmat2017} and the first derivation of FRW equations with torsion was presented in \cite{tsamPRD}.

The present paper aims to study torsion effects in FRW background for late time expansion of the universe, particularly the possibility to explain the recent accelerated phase of expansion as a consequence of torsion effects. It is assumed four different types of torsion function, parameterized by one, two and three free parameters. Constraints with observational data allows to fix the free parameters and compare the known parameters with the ones obtained from standard cosmological model, namely, the $\Lambda$CDM model parameters obtained from last Planck satellite observations \cite{Aghanim2018}.

The paper is organised as follows. Section 2 presents the main equations of Friedmann cosmology with torsion, based on \cite{Kranas2019}. In Section 3, the constraints from observational data are obtained for four different torsion functions. Section 4 analyses the torsion function and deceleration parameter evolution for the best function obtained in previous section. Conclusion is left to Section 5.

\section{Friedmann cosmology with torsion}
\label{sec:torsion}

We follow the same notation from \cite{Kranas2019}. The standard  Einstein equation of gravitation maintain its original form in terms of Ricci tensor, Ricci scalar and energy momentum tensor,
\begin{equation}
    R_{\mu\nu} - \frac{1}{2}R g_{\mu\nu}=\kappa T_{\mu\nu}\,,\label{Rmunu}
\end{equation}
    with $\kappa = {8\pi G}$, however in a space-time with torsion the affine connection is endowed with an antisymmetric part, namely $\Gamma^\alpha_{~~~\mu\nu}=\Tilde{\Gamma}^\alpha_{~~~\mu\nu}+K^\alpha_{~~~\mu\nu}$, where $\Tilde{\Gamma}^\alpha_{~~~\mu\nu}$ defines the symmetric Christoffel symbols and $K^\alpha_{~~~\mu\nu}$ defines the contorsion tensor,
    \begin{equation}
        K^\alpha_{~~~\mu\nu} = S^\alpha_{~~~\mu\nu} + S_{\mu\nu}^{~~~\alpha} + S_{\nu\mu}^{~~~\alpha}\,\label{K}
    \end{equation}
    written in terms of the torsion tensor $S^\alpha_{~~~\mu\nu}$, which is antisymmetric in its covariant indices, $S^\alpha_{~~~\mu\nu} = - S^\alpha_{~~~\nu\mu}$. In general case the energy momentum tensor is coupled to $S^\alpha_{~~~\mu\nu}$  by means of the Cartan field equations,
    \begin{equation}
        S_{\alpha\mu\nu} = -\frac{1}{4}\kappa (2s_{\mu\nu\alpha}+g_{\nu\alpha}s_\mu - g_{\alpha\mu}s_\nu)\,, \label{Sss}
    \end{equation}
    where $s_{\alpha\mu\nu}$  and $s_\alpha=s^\mu_{~~\alpha\mu}$ are the tensor and vector spin of matter, respectively. Physically, torsion provide a link between the
    spacetime geometry and the intrinsic angular momentum of the matter \cite{Pasmat2017}. With the presence of torsion terms into Eq. (\ref{Rmunu}), it is known as the Einstein-Cartan equation of gravitation. 
    
    In a homogeneous and isotropic Friedmann background, the torsion tensor and the associated vector are \cite{Kranas2019}
    \begin{equation}
     S_{\alpha\mu\nu} = \phi (h_{\alpha\mu} u_\nu - h_{\alpha\nu}u_\mu)   \hspace{1cm}  S_\alpha = -3\phi u_\alpha\,, \label{SS}
    \end{equation}
    where $\phi=\phi(t)$ is an unique time dependent function representing torsion contribution due to homogeneity of space,  $h_{\mu\nu}$ is a projection tensor, symmetric and orthogonal to the 4-vector velocity $u_\mu$.
    
    In terms of the torsion field $\phi(t)$, the Friedmann equations are \cite{Kranas2019}:
    \begin{equation}
        \bigg(\frac{\dot{a}}{a}\bigg)^2 =\frac{8\pi G}{3}\rho - \frac{k}{a^2}-4\phi^2 - 4\bigg(\frac{\dot{a}}{a}\bigg)\phi\,,\label{H2}
    \end{equation}
    \begin{equation}
         \frac{\ddot{a}}{a} =-\frac{4\pi G}{3}(\rho+3p) - 2\dot{\phi} - 2\bigg(\frac{\dot{a}}{a}\bigg)\phi\,,\label{Hd2}
    \end{equation}
    where $k$ is the curvature parameter, $\rho$ and $p$ are the energy density and pressure of matter. Together the Friedmann equations, for a barotropic matter satisfying $p=\omega \rho$, the continuity equation reads \cite{Kranas2019}
    \begin{equation}
        \dot{\rho}+3(1+\omega)H\rho + 2(1+3\omega)\phi \rho = 0\,,\label{rhodot}
    \end{equation}
    with $H=\dot{a}/a$, whose solution with $\omega$ constant is
\begin{equation}
    \rho = \rho_0\bigg(\frac{a_0}{a}\bigg)^{3(1+\omega)} \textrm{e}^{-2(1+3\omega)\int_{t_i}^{t}\phi(t) dt}
\end{equation}
where the subscript $0$ denotes present values and $t_i$ some initial time. We see that torsion alters the energy density evolution of standard matter through the exponential term. 


In order to better understand the influence of torsion function into recent accelerated phase of expansion of the universe, we look for the deceleration parameter, which can be written as
\begin{equation}
    q =\frac{4\pi G}{3H^2}(\rho+3p) + 2\frac{\dot{\phi}}{H^2} + 2\frac{\phi}{H}\,.\label{qt}
\end{equation}
For a constant and negative $\phi$ for instance, torsion tends to accelerate the expansion. For a flat and empty space ($k=\rho=0$), Eq. (\ref{H2}) leads to $\phi(t)=-H(t)/2$, which suggest a $H$ dependence to the torsion function. However, since the physical source of torsion is the spin of matter, a torsion function dependent on the matter density is also a much more realistic choice. The above discussions and dimensional arguments will guide us in the next section in order to build some torsion functions and use them to compare with observational constraints, constraining the free parameters.

\section{Constraints from observational data}
\label{sec:constraint}

In order to study the possibility of dark matter and dark energy being driven by torsion effects in cosmological evolution, let us analyse the constraints imposed by observational data in four different models of torsion fields together ordinary matter contribution. Cases I and II below are just phenomenological assumptions for the torsion function, the first a constant function and the second evolving with $H(t)$. Cases III and IV are more realistic once they are explicitly dependent on the matter energy density, the real sources of spin in the universe.

The data used here were 51 $H(z)$ data from Maga\~na {\it et al.} \cite{Magana2018} and 1048 SNe Ia data from Pantheon compilation \cite{Scolnic2018}.

In all analyses here, we have written a $\chi^2$ function for parameters, with the likelihood given by ${\mathcal L}\propto e^{-\chi^2/2}$. The $\chi^2$ function for $H(z)$ data is given by
\begin{equation}
\chi^2_H = \sum_{i = 1}^{51}\frac{{\left[ H_{obs,i} - H(z_i,\mathbf{s})\right] }^{2}}{\sigma^{2}_{H_i,obs}} ,
\label{chi2H}
\end{equation}
where $\mathbf{s}$ is the parameter vector. For Pantheon, instead, we included systematic errors, thus we had to deal with the full covariance matrix. In this case, the $\chi^2$ is given by
\begin{equation}
\chi^2_{SN} = \left[\bm{m}_{obs}-{\bm m}(z,\bm{s})\right]^T\bm{C}^{-1}\left[\bm{m}_{obs}-{\bm m}(z,\bm{s})\right]
\label{chi2SN}
\end{equation}
where $\bm{C}$, $\bm{m}_{obs}$ and $\bm{m}$ are covariance matrix, observed apparent magnitude vector and model apparent magnitude, respectively. We have assumed flat priors for all parameters and have sampled the posteriors with the so called Affine Invariant Monte Carlo Markov Chain (MCMC) Ensemble Sampler by \cite{GoodWeare}, which was implemented in {\sffamily Python} language with the {\sffamily emcee} software by \cite{ForemanMackey13}. In order to plot all the constraints on each model, we have used the freely available software {\sffamily getdist}\footnote{{\sffamily getdist} is part of the great MCMC sampler, {\sffamily COSMOMC} \cite{cosmomc}.}, in its {\sffamily Python} version.

\subsection{Case I: $\phi(t)=\phi_0 = -\alpha H_0$}

For this simplest case of a constant torsion field, as already discussed by \cite{Kranas2019}, with $\alpha$ a dimensionless constant to be determined\footnote{The presence of $H_0$ warrants the correct dimension for the torsion term.},  we write the Friedmann equation (\ref{H2}) as:
\begin{equation}
    H^2=\frac{8\pi G}{3}\rho_m - \frac{k}{a^2}-4\alpha H_0 H - 4 \alpha^2 H_0^2\,, \label{H2case1}
\end{equation}
where $\rho_{m}$ is the matter density parameter obtained as a solution of (\ref{rhodot}) with $\omega =0$, namely $\rho_m = \rho_{0m}\big({a_0}/{a}\big)^{3} \textrm{e}^{2\alpha H_0 t}$, where $\rho_{0m}$ represents the present day matter energy density. Analytic solution of (\ref{H2case1}) exists just for spatially flat $(k=0)$ background, however a numeric treatment can be done in general case and the parameters $\alpha$, 
$\Omega_{m}$ and $H_0$ can be constrained with observational data\footnote{Here $\Omega_{m}=\frac{\rho_{0m}}{\rho_{0c}}$ as usual, and $\rho_{0c}=\frac{3H_0^2}{8\pi G}$ is the critical density.}.

\begin{figure}[t] 
\centering
\includegraphics[width=0.8\linewidth]{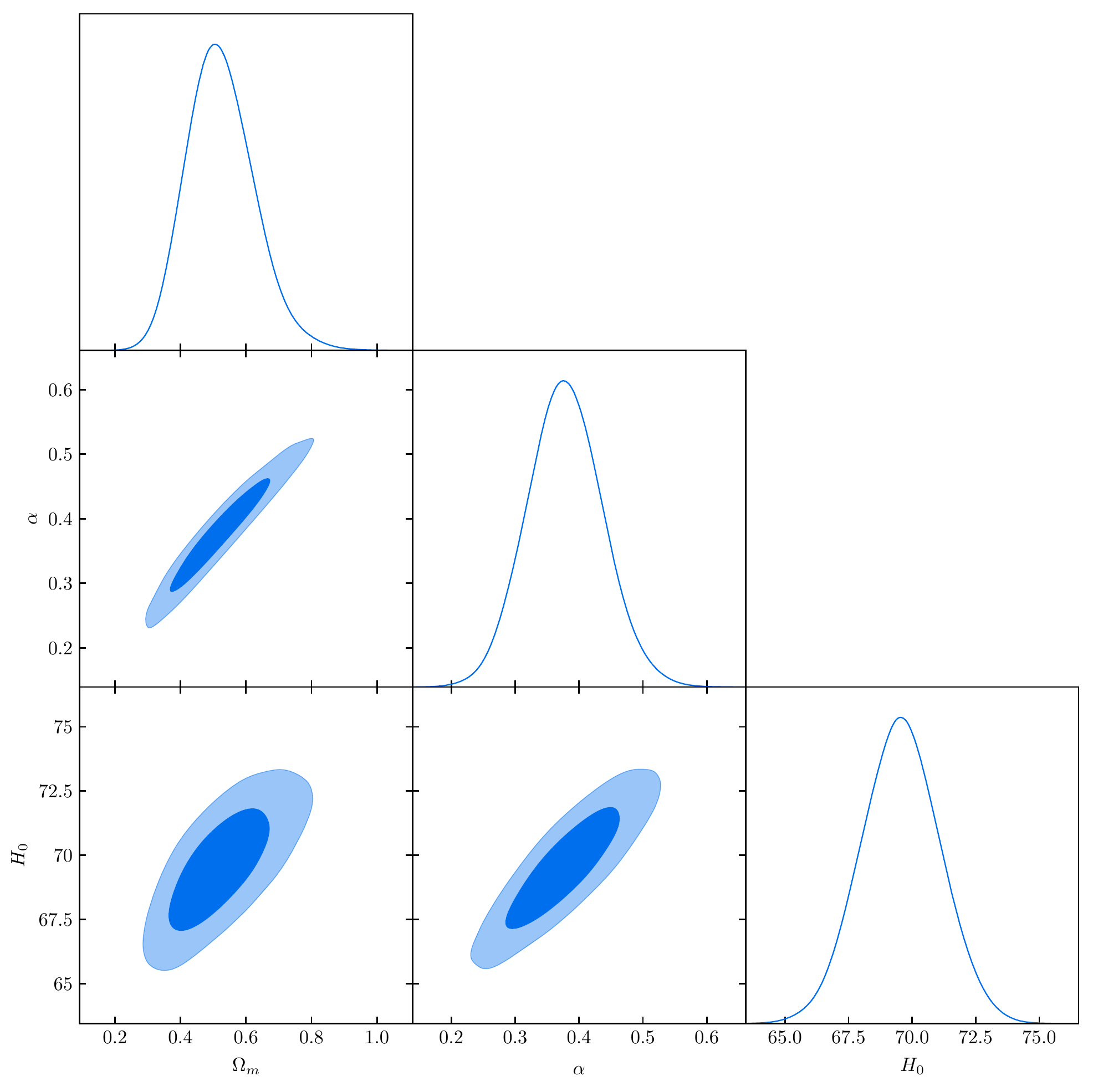}
\caption{SNe Ia$+H(z)$ constraints in Case I.}
\label{contI}
\end{figure}

Figure \ref{contI} shows the $1 \sigma$ (68.3\% c.l.) and $2 \sigma$ (95.4\% c.l.) contours for $\Omega_m$, $\alpha$ and $H_0$ parameters obtained with $H(z)$ and SNe Ia observational data. Table 1 presents the mean values of the parameters with $95\%$ c.l. constraints. For this model we see that $\Omega_m$ is just marginally compatible at $2\sigma$ with the last results for $\Lambda$CDM model from the Planck collaboration on the cosmological parameters\footnote{From \cite{Aghanim2018}, $\Omega_m = 0.315\pm 0.007$ for matter density and $H_0=(67.4\pm 0.5)$ km/s/Mpc.} \cite{Aghanim2018}, while $H_0$ is compatible at $1\sigma$.

\begin{table}[!h]
    \centering
    \begin{tabular} { l  c}
\hline
 Parameter &  95\% limits\\
\hline
{\boldmath$\Omega_m       $} & $0.52^{+0.21}_{-0.20}      $\\

{\boldmath$\alpha         $} & $0.38^{+0.12}_{-0.11}      $\\

{\boldmath$H_0            $} & $69.6^{+3.1}_{-3.1}        $\\
\hline
\end{tabular}
    \caption{Mean values of the free parameters and 95\% c.l. constraints for Case I.}
    \label{tab:my_label}
\end{table}

\subsection{Case II: $\phi(t)=-\alpha H(t)$}

For this case, the analytic solution of (\ref{rhodot}) is
\begin{equation}
    \rho_m=\rho_{0m}\bigg(\frac{a}{a_0}\bigg)^{-3+2\alpha}
\end{equation}
and the Friedmann equation (\ref{H2}) turns:
\begin{equation}
    H^2=\frac{8\pi G}{3}\rho_m - \frac{k}{a^2}+4\alpha  H^2 - 4 \alpha^2 H^2\,, \label{H2case2}
\end{equation}
In terms of the density parameters, Eq. (\ref{H2case2}) is:
\begin{equation}
    \frac{H}{H_0}= \sqrt{\frac{\Omega_{m}(1+z)^{3-2\alpha}+\Omega_k(1+z)^2}{1-4\alpha+4\alpha^2}}\, , \label{Hcase2}
\end{equation}
where\footnote{$\Omega_k\equiv -\frac{k}{a_0^2 H_0^2}$ is the curvature parameter} $\Omega_k=1-\Omega_{m} -4\alpha+4\alpha^2$ and the redshift is introduced by $(1+z)=a_0/a$.

\begin{figure}[t] 
\centering
\includegraphics[scale=0.4]{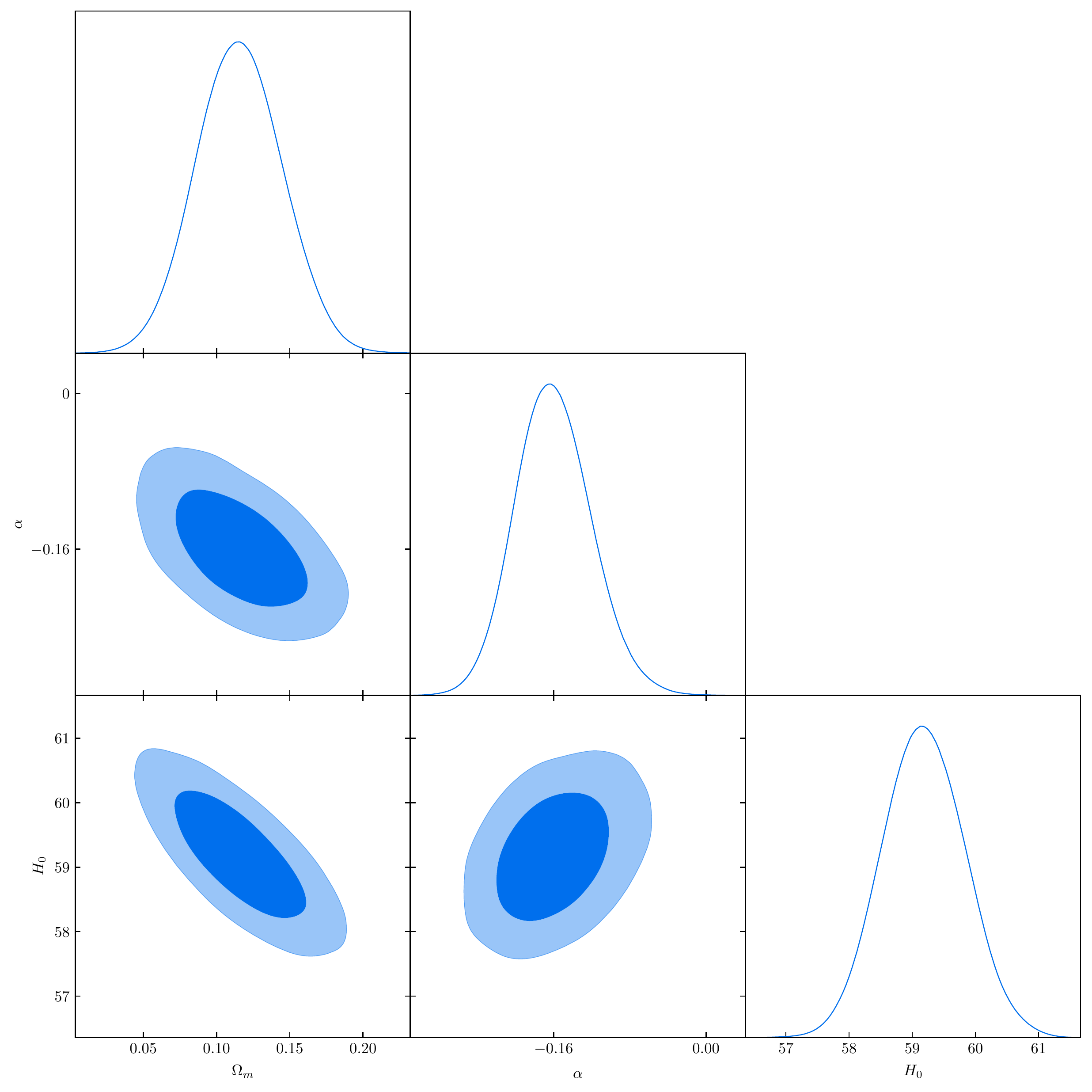}
\caption{SNe Ia + $H(z)$ constraints for Case II.}
\label{contII}
\end{figure}

Figure \ref{contII} shows the constraints for $\Omega_m$, $\alpha$ and $H_0$  at $1 \sigma$ and $2 \sigma$ contours for 
$H(z)$ and SNe Ia observational data. Table \ref{tabII} presents the mean values of the parameters with $95\%$ c.l. constraints. For this model we see that both $\Omega_m$ and $H_0$ are very small, not compatible with the $\Lambda$CDM model even at $2\sigma$.
\begin{table}[!h]
    \centering
    \begin{tabular} { l  c}
\hline
 Parameter &  95\% limits\\
\hline
{\boldmath$\Omega_m       $} & $0.116^{+0.059}_{-0.057}   $\\

{\boldmath$\alpha         $} & $-0.160^{+0.084}_{-0.076}  $\\

{\boldmath$H_0            $} & $59.2\pm1.3$\\
\hline
\end{tabular}
    \caption{Mean values of the free parameters and 95\% c.l. constraints for Case II.}
    \label{tabII}
\end{table}

\subsection{Case III: $\phi(t)=- \alpha H(t) \bigg(\frac{\rho_{m}(t)}{\rho_{0c}}\bigg)^n$}

This general case is much more interesting, since that the torsion function is proportional to matter density $\rho_m$ and it is expected that torsion contribution comes from spin of ordinary matter. Also, for this case it is easy to verify that a solution of Eq. (\ref{rhodot}) for the energy density with $\omega=0$ is
\begin{equation}
    \rho_m(a) = \rho_{0c}\frac{3^{1/n}}{\big(2\alpha + 3C_1 (a/a_0)^{3n}\big)^{1/n}}\,,\label{rho03}
\end{equation}
where $C_1$ is a integration constant. In order to 
have $\rho_m(a_0)=\rho_{m0}$, we set $C_1=-\frac{2}{3}\alpha + \Omega_{m}^{-n}$.

The Friedmann equation (\ref{H2}) is:
\begin{equation}
    H^2=\frac{8\pi G}{3}\rho_m - \frac{k}{a^2}+4 \alpha H^2 \bigg(\frac{\rho_{m}}{\rho_{0c}}\bigg)^n - 4 \alpha^2 H^2\bigg(\frac{\rho_{m}}{\rho_{0c}}\bigg)^{2n}\,, \label{H2case3}
\end{equation}
In terms of the density parameters Eq. (\ref{H2case3}) is:
\begin{equation}
    \frac{H}{H_0}=\frac{(3-2\alpha \Omega_{m}^n)+2\alpha \Omega_{m}^n(1+z)^{3n}}{(3-2\alpha \Omega_{m}^n)-4\alpha \Omega_{m}^n(1+z)^{3n}}\sqrt{\frac{  3^{1/n}\Omega_{m}}{\left[2\alpha\Omega_{m}^n +\frac{(3-2\alpha \Omega_{m}^n)}{(1+z)^{3n}}\right]^{1/n}} +\Omega_k (1+z)^2}\, , \label{Hcase3}
\end{equation}
where $\Omega_k=(1-2\alpha\Omega_{m}^n)^2 - \Omega_{m}$. 

\begin{figure}[t] 
\centering
\includegraphics[width=.8\linewidth]{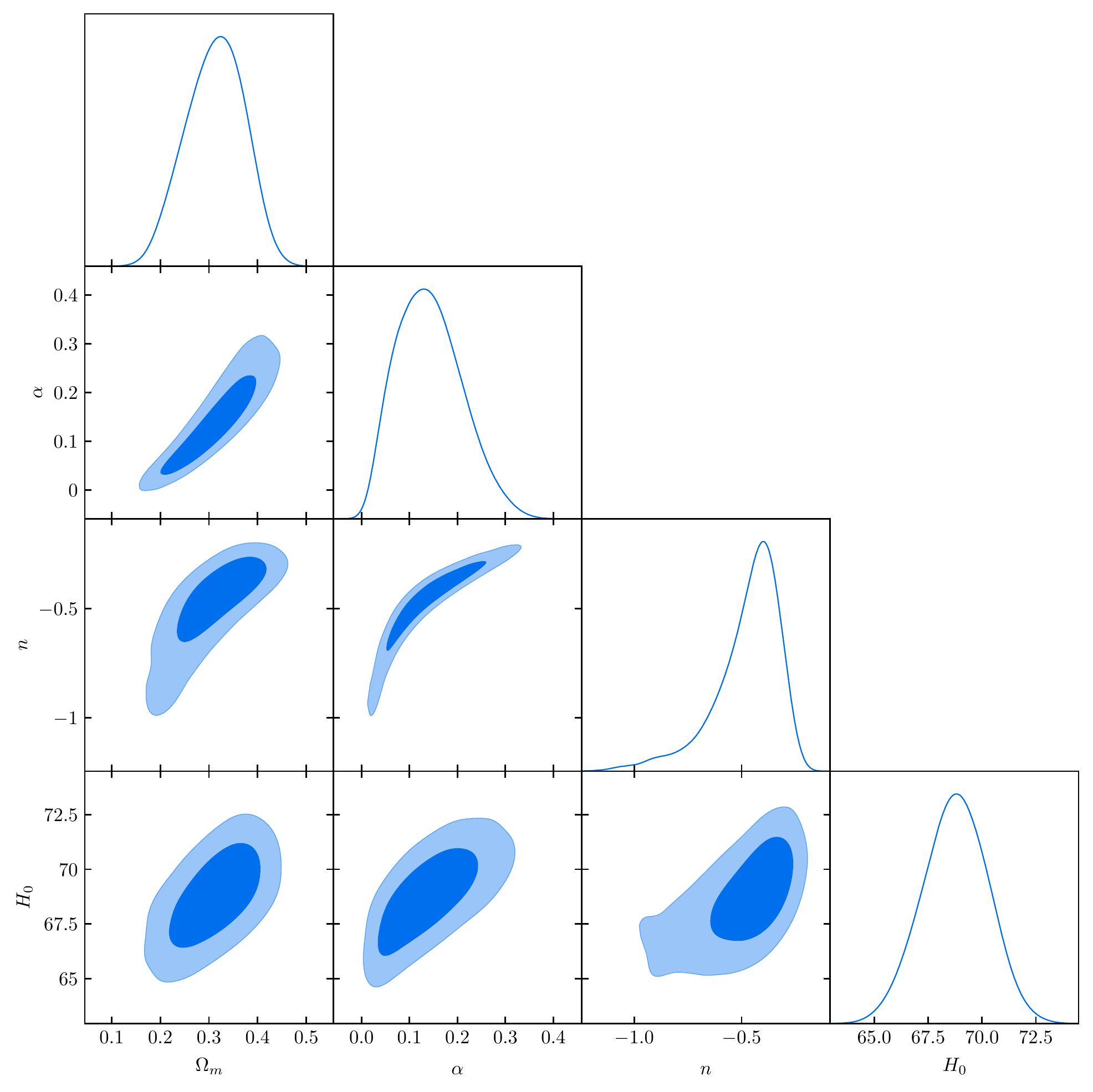}
\caption{Constraints from SNe Ia$+H(z)$ for Case III.}
\label{contIII}
\end{figure}

Figure \ref{contIII} shows the constraints for $\Omega_m$, $n$, $\alpha$ and $H_0$ at $1 \sigma$ and $2 \sigma$ contours for $H(z)$ and SNe Ia observational data. Table \ref{tabIII} presents the mean values of the parameters with $95\%$ c.l.. We see that both $\Omega_m$ and $H_0$ are in very good agreement to the $\Lambda$CDM model at $1\sigma$, with a small positive $\alpha$ value and a negative $n$ value. With such parameters the model is totally compatible with the recent cosmic acceleration, with the dark energy component being represented by torsion function $\phi(t)$.
\begin{table}[!h]
    \centering
    \begin{tabular} { l  c}
\hline
 Parameter &  95\% limits\\
\hline
{\boldmath$\Omega_m       $} & $0.31^{+0.11}_{-0.12}      $\\

{\boldmath$\alpha         $} & $0.14^{+0.14}_{-0.12}      $\\

{\boldmath$n              $} & $-0.47^{+0.26}_{-0.36}     $\\

{\boldmath$H_0            $} & $68.8^{+3.0}_{-3.1}        $\\
\hline
\end{tabular}
    \caption{Mean values of the free parameters and 95\% c.l. constraints for Case III.}
    \label{tabIII}
\end{table}

\subsection{Case IV: $\phi(t)=- \alpha H_0\bigg(\frac{H_0}{H(t)}\bigg)^m \bigg(\frac{\rho_{m}(t)}{\rho_{0c}}\bigg)^n$}

For this general case there is no analytic solution for the energy density $\rho_m$ and one must resort to numerical methods. Due to this model having many free parameters, we choose to work with the spatially flat case ($k=0$), which is favoured by inflation and recent CMB observations.

The Friedmann equation (\ref{H2}) for a spatially flat Universe is:
\begin{equation}
    H^2=\frac{8\pi G}{3}\rho_m +4 \alpha H_0^{m+1}H^{-m+1} \bigg(\frac{\rho_{m}}{\rho_{0c}}\bigg)^n - 4 \alpha^2 H_0^{m+2}H^{-2m}\bigg(\frac{\rho_{m}}{\rho_{0c}}\bigg)^{2n}\,, \label{H2case4}
\end{equation}

{Due to spatial flatness, there is a relation among the free parameters of this model, which can be obtained from the Friedmann equation, as:}
\begin{equation}
\Omega_m=\left(1+\frac{2\phi_0}{H_0}\right)^2=(1-2\alpha\Omega_m^n)^2
\label{wmphi}
\end{equation}
{where the last equality holds only for Case IV. As we can see from this relation, due to nonlinearity, there is not a unique solution for parameters $\alpha$ or $n$. So, for this analysis, we choose to work with the free parameter $\varphi_0\equiv\frac{\phi_0}{H_0}$, we find the constraints over $\varphi_0$ and then we use it to obtain $\Omega_m$ by using (\ref{wmphi}). We choose a flat prior $\varphi_0\in[-1,0]$, which yields a flat prior $\Omega_m\in[0,1]$.}

\begin{figure}[t] 
\centering
\includegraphics[width=\linewidth]{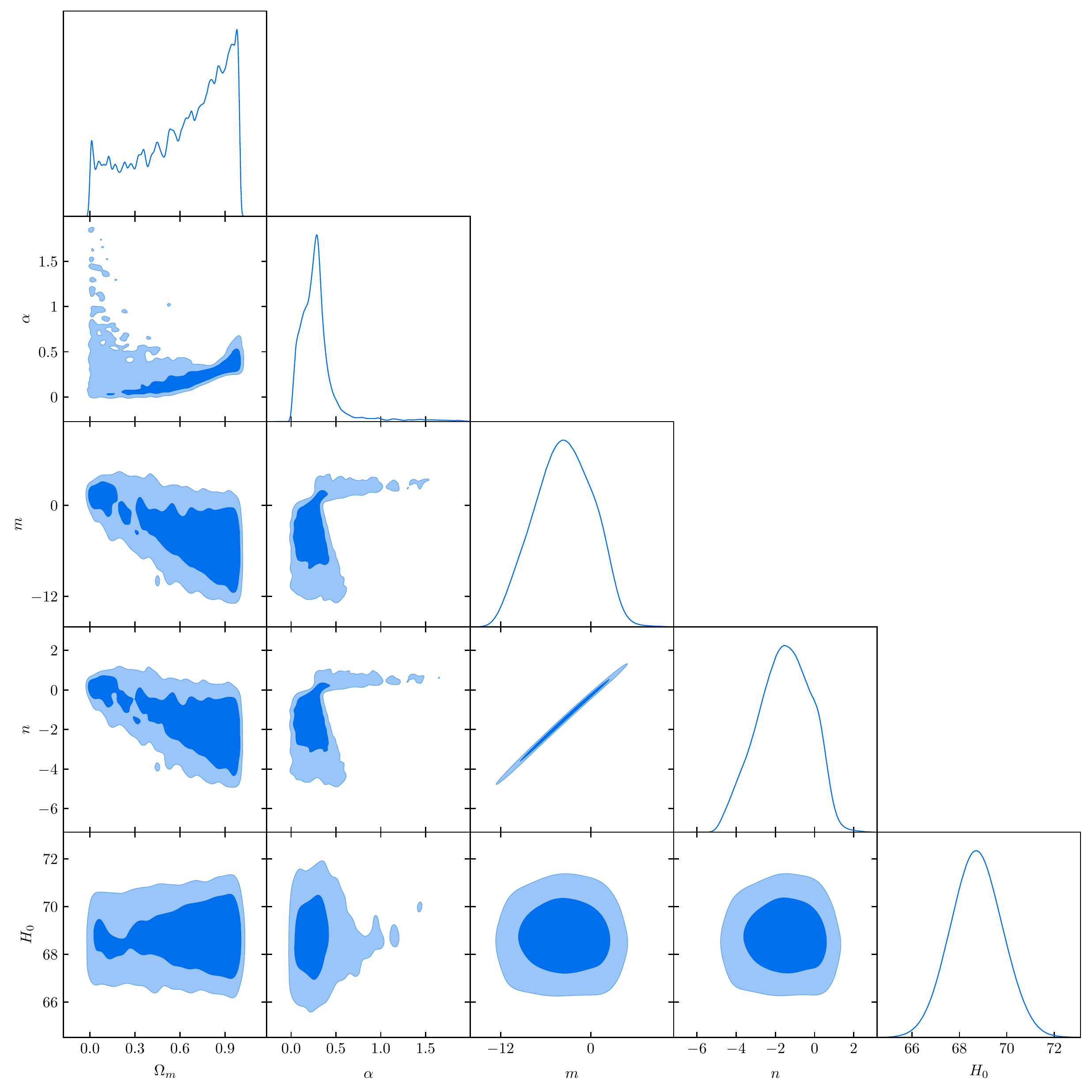}
\caption{Constraints from SNe Ia$+H(z)$ for Case IV.}
\label{contIV}
\end{figure}

Figure \ref{contIV} shows the constraints for $\Omega_m$, $n$, $m$, $\alpha$ and $H_0$ at $1 \sigma$ and $2 \sigma$ contours for $H(z)$ and SNe Ia observational data. Table \ref{tabIV} presents the mean values of the parameters with $95\%$ c.l.. {We see that $\Omega_m$ and $H_0$ are compatible with the $\Lambda$CDM model values. Also, one can see that $\Omega_m$ is poorly constrained by this   analysis, due to having too many free parameters.}
\begin{table}[!h]
    \centering
    \begin{tabular} { l  c}
\hline
 Parameter &  95\% limits\\
\hline
$\Omega_m                  $ & $0.62^{+0.39}_{-0.62}      $\\
{\boldmath$\alpha         $} & $0.30^{+0.51}_{-0.31}      $\\
{\boldmath$m              $} & $-3.6^{+6.9}_{-7.6}        $\\
{\boldmath$n              $} & $-1.5^{+2.4}_{-2.7}        $\\
{\boldmath$H_0            $} & $68.7\pm2.2        $\\
\hline
\end{tabular}
    \caption{Mean values of the free parameters and 95\% c.l. constraints for Case IV.}
    \label{tabIV}
\end{table}

\newpage

{
Now we can compare the mean values for $\Omega_m$ and $H_0$ cosmological parameters obtained for the above four cases with the ones from $\Lambda$CDM model, together with the statistical parameters as $\chi^2_{min}$, AIC and BIC \cite{Akaike74,Schwarz78,JesusEtAl16}. We choose to compare the torsion models with both concordance models, namely spatially flat $\Lambda$CDM model, as well as with the O$\Lambda$CDM model, which allows for nonzero spatial curvature, once that only Case IV is spatially flat. Table \ref{tabBIC} present the parameters at 95\% c.l. for the combined analysis $H(z)$+Pantheon.}
\begin{table}
\centering
\begin{tabular}{lcccccc}
\hline
 & Case I & Case II & Case III  & Case IV & O$\Lambda$CDM & $\Lambda$CDM\\
\hline
{\boldmath$\Omega_m$} & $0.52^{+0.21}_{-0.20}$ & $0.116^{+0.059}_{-0.057}$ & $0.31^{+0.11}_{-0.12}$ & $0.62^{+0.39}_{-0.62}$ & $0.238^{+0.056}_{-0.057}$ & $0.278^{+0.031}_{-0.028}$\\
{\boldmath$H_0$} & $69.6\pm3.1$ & $59.2\pm1.3$ & $68.8^{+3.0}_{-3.1}$ & $68.7\pm2.2$ & $68.0^{+2.7}_{-2.6}$ & $69.57^{+0.99}_{-0.98}$\\
{\boldmath$\chi^2_{min}$} & 1068.23 & 1097.90 & 1053.28 & 1054.20 & 1055.15 & 1057.74\\
{\boldmath$\chi^2_\nu$} & 0.9747 & 1.0017 & 0.9619 & 0.9627 & 0.9627 & 0.9642\\
{\boldmath$p$} & 3 & 3 & 4 & 4 & 3 & 2\\
AIC & 1074.23 & 1103.90 & 1061.28 & 1062.20 & 1061.15 & 1061.74\\
BIC & 1089.24 & 1118.91 & 1081.29 & 1082.21 & 1076.16 & 1071.74\\
\hline
\end{tabular}
\caption{Common cosmological parameters ($\Omega_m$ and $H_0$) from Cases I, II, III, IV and from $\Lambda$CDM, O$\Lambda$CDM models at 95\% c.l. and respective statistical parameters. $\chi^2_\nu=\chi^2_{min}/(n-p)$, where $n$ is number of data and $p$ is number of free parameters. $\mathrm{AIC}=\chi^2_{min}+2p$, $\mathrm{BIC}=\chi^2_{min}+p\ln N$.}
\label{tabBIC}       
\end{table}

{As we can see on this Table, AIC favours O$\Lambda$CDM and Case III over the concordance flat $\Lambda$CDM. But it is a known fact \cite{JesusEtAl16} that AIC does not penalize enough the excess of parameters, and BIC is in general most accepted, being an approximation of the Bayesian Evidence. According to BIC, the flat $\Lambda$CDM model and O$\Lambda$CDM are favoured by this analysis. Case III is the torsion model with the lowest BIC, although significantly above $\Lambda$CDM.}

\section{Torsion evolution and transition redshift}

In order to better reproduce the standard model constraints and obtain a cosmic acceleration in agreement with the latest observational data (see footnote 3), Case III above is the better one, with both $\Omega_m$ and $H_0$ compatible within $1\sigma$ c.l. {and with lower $\chi^2$ and BIC parameters.}

\begin{figure}[t] 
\centering
\includegraphics[scale=0.5]{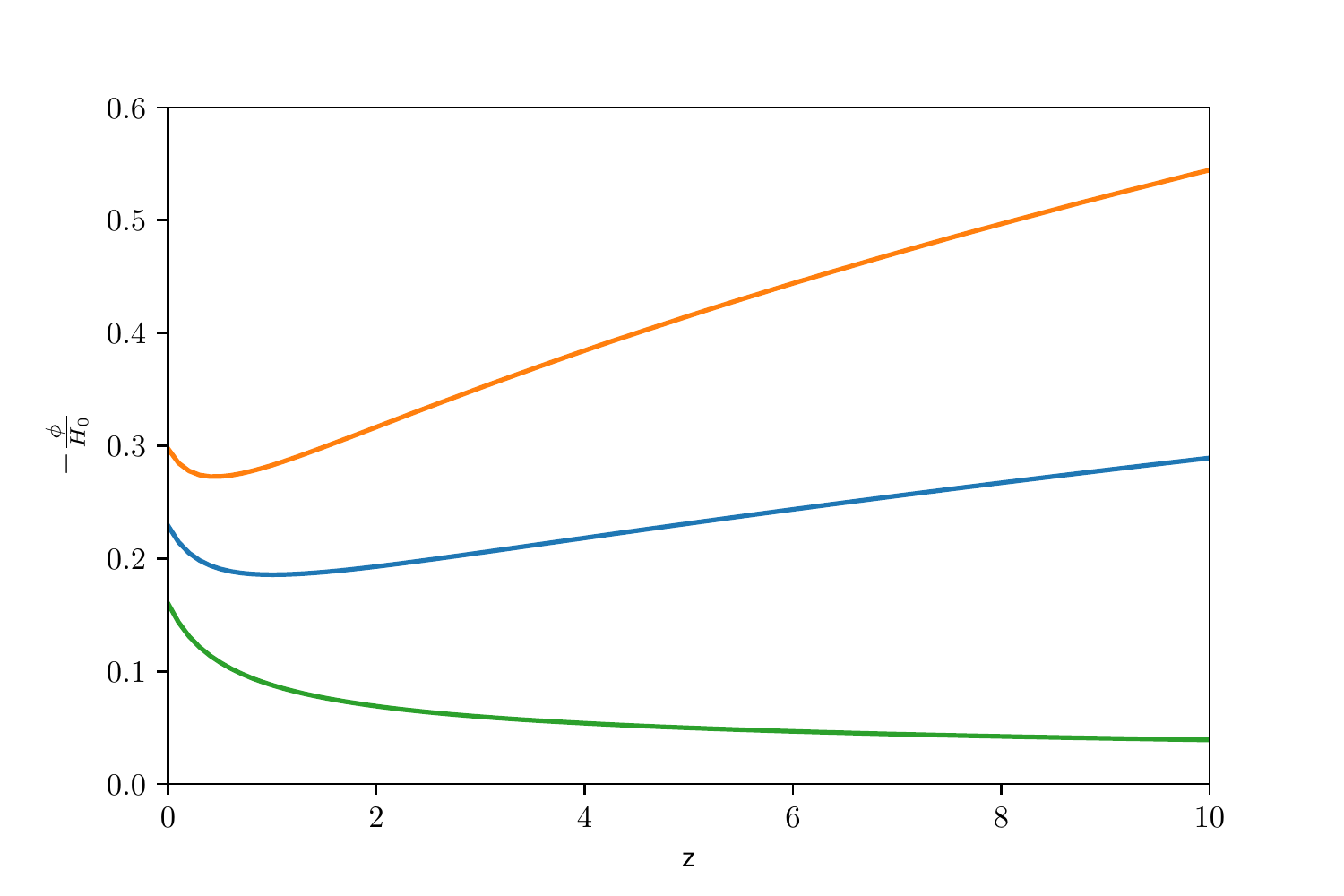}
\caption{Evolution of $\phi(z)$ for the mean values of parameters (blue line) and for $1\sigma$ c.l. (orange and green lines).}
\label{phiz}
\end{figure}

For this case it is interesting to analyse the evolution of torsion function and the transition redshift. From (\ref{rho03}) and (\ref{H2case3}), 
we have obtained the mean $\phi(z)$ from the parameters MCMC chains, jointly with its variance. The evolution of the torsion function is shown in Figure \ref{phiz}, for the mean $\phi(z)$ (blue line) and for $1\sigma$ c.l. (orange and green lines). The behaviour of the torsion function on the past is strongly dependent on the values of parameters, specifically on the $n$ parameter. At present, the behaviour is similar in all cases, showing an increase on the absolute value of torsion function just in recent times, which coincides with the late time acceleration phase of expansion of universe. In this sense, torsion function makes the role of a dark energy acting during the whole history of the universe. In the past the matter energy density dominates over the torsion contribution and today is the torsion function that dominates, driving the acceleration. {A similar observation was made in a recent work within the context of a scalar-tensor theory with torsion \cite{Cid:2017wtf}, where the authors showed that an effective torsional fluid plays an important role in recent acceleration phase of expansion of universe but becomes subdominant in the past, where a pressureless matter component dominates at redshift around 200.}

\begin{figure}[t] 
\centering
\includegraphics[scale=0.5]{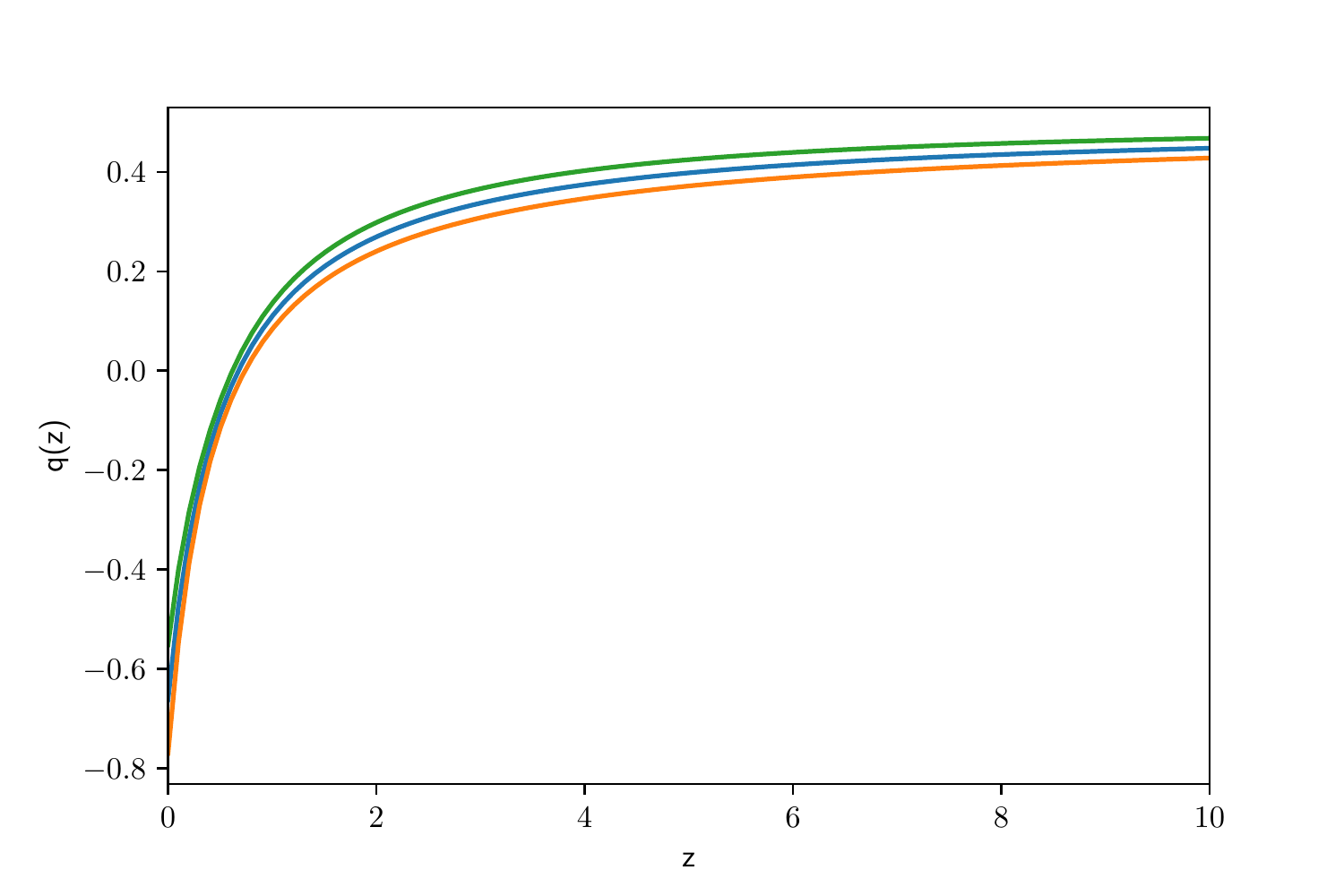}
\caption{Evolution of $q(z)$ for the mean values of parameters (blue line) and for $1\sigma$ c.l. (orange and green lines).}
\label{qz}
\end{figure}

The behaviour of the deceleration parameter is better to understand the recent evolution of the universe and is presented on Figure \ref{qz}. For larger $z$ values the deceleration parameter seems to converge to $0.5$, a value characteristic of a matter dominated universe, as expected from standard model. As seen above, for $z\lesssim 1$ the torsion function start to increase and a transition to accelerated phase occurs, dominated by torsion term. The transition redshift $z_t$ occurs at about $z_t=0.65$, in good agreement to standard model.

\section{Conclusion}\label{sec 6}

We have analysed a Friedmann like universe with the contribution of a torsion function in Einstein-Cartan cosmology. The torsion function is represented by $\phi(t)$, and for four different types of function written in terms of one, two and three free parameters we have studied the cosmic evolution and constrained the free parameters with observational data from $H(z)$ and SN Ia. 

From the four different functions, Case III presents a very good agreement to observational data, in the sense that both matter energy density parameter and $H_0$ are completely compatible with the results for $\Lambda$CDM model obtained from last Planck mission observations. {Statistical parameters as BIC and $\chi^2$ also favours Case III, as pointed out in Table 5, comparing the four models with standard $\Lambda$CDM model.}

The effect of torsion function is to act as a dark energy fluid at late time, correctly explaining the present accelerated phase of expansion of the universe. {A model-independent
result concerning Einstein-Cartan gravity was obtained in \cite{ivanov2016}, showing that torsion can be responsible for the vacuum energy
density or the cosmological constant, exactly as dark energy density in the universe}. 

Finally, the deceleration parameter obtained for the model furnish a desirable transition to accelerated phase at about $z_t=0.65$, coming from a matter dominated phase in the past, as occurs for standard model of cosmology.

\begin{acknowledgments}
This study was financed in part by the Coordena\c{c}\~ao de Aperfei\c{c}oamento de Pessoal de N\'ivel Superior - Brasil (CAPES) - Finance Code 001. SHP would like to thank CNPq - Conselho Nacional de Desenvolvimento Cient\'ifico e Tecnol\'ogico, Brazilian research agency, for financial support, grants number 303583/2018-5 and 400924/2016-1. RCL would like to thank CAPES. JFJ is supported by Funda\c{c}\~ao de Amparo \`a Pesquisa do Estado de S\~ao Paulo - FAPESP (Process no. 2017/05859-0).
\end{acknowledgments}


\end{document}